\documentclass[preprint]{elsarticle}

\usepackage{hyperref}

\begin{document}

\begin{frontmatter}

\title{Humans make best use of social heuristics when confronting hard problems in large groups}

\author{Federica Stefanelli$^1$, Enrico Imbimbo$^1$, Daniele Vilone$^{2,3}$, Franco Bagnoli$^4$, Zoran Levnaji\'c$^5$, Andrea Guazzini$^1$}
\address{\ \\
$^1$Department of Education and Psychology, University of Florence, Florence, Italy \\
\ \\
$^2$Laboratory of Agent Based Social Simulation, Institute of Cognitive Science and Technology, National Research Council (CNR), Rome, Italy; \\
\ \\
$^3$Grupo Interdisciplinar de Sistemas Complejos (GISC), Departamento de Matem\'aticas, Universidad Carlos III de Madrid, Spain \\
\ \\
$^4$Department of Physics and Astronomy \&
Center for Study of Complex Dynamics, University of Florence and INFN, Florence, Italy \\
\ \\
$^5$Faculty of Information Studies in Novo mesto, Novo Mesto, Slovenia
}


\begin{abstract}
We report the results of a game-theoretic experiment with human players who solve the problems of increasing complexity by cooperating in groups of increasing size. Our experimental environment is set up to make it complicated for players to use rational calculation for making the cooperative decisions. This environment is directly translated into a computer simulation, from which we extract the collaboration strategy that leads to the maximal attainable score. Based on this, we measure the error that players make when estimating the benefits of collaboration, and find that humans massively underestimate these benefits when facing easy problems or working alone or in small groups. In contrast, when confronting hard problems or collaborating in large groups, humans accurately judge the best level of collaboration and easily achieve the maximal score. Our findings are independent on groups' composition and players' personal traits. We interpret them as varying degrees of usefulness of social heuristics, which seems to depend on the size of the involved group and the complexity of the situation.
\end{abstract}


\end{frontmatter}


\section{Introduction}

Cooperation allows to solve problems that are too complex for anyone to solve individually. It allows to concentrate the efforts of many individuals on tasks that do not give immediate reward to none of them. It is one of the key factors that lead humans to become possibly the most cooperative species in nature, able to collaborate at (almost) every level of our society \citep{nowak2011supercooperators,perc2017statistical}. 

But the benefits that cooperative behavior brings come with a price for The Individual. This apparently odd dualism and the social mechanisms behind it has been the focus of research attention for decades~\citep{helbing2015saving}. Specifically,
research identified several factors influencing pro-social behavior, including direct \citep{rand2009direct,delton2011evolution} and indirect \citep{panchanathan2004indirect,nowak2005evolution} reciprocity, multilevel cooperation \citep{traulsen2006evolution, szolnoki2009emergence}, kin-selection \citep{lieberman2007architecture}, and social network structure \citep{ohtsuki2006simple,rand2011dynamic}. These mechanisms seem to be responsible for transforming the cooperation into an evolutionary stable strategy and spreading it within and among human societies \citep{nowak2006}. 
In fact, we routinely rely on various collaboration strategies to solve problems of different nature in our daily lives. This instinctive drive for cooperation has long been a topic of interest \citep{bowles2003origins,jordan2016uncalculating,mann2017optimal}, since understanding the factors behind this drive could promote and regulate successful collaboration among human beings. These processes are often studied in form of ``crowdsourcing'' or ``wisdom of the crowds'' \citep{howe2006rise,surowiecki2007wisdom,szolnoki2012wisdom,guazzini2015,prelec2017solution}. In particular, it was found that cooperative organized actions can lead to original ideas and solutions of notoriously hard problems. These include improving medical diagnostics \citep{kurvers2016boosting}, designing RNA molecules \citep{lee2014rna}, computing crystal \citep{fdetermining} or predicting protein structures \citep{cooper2010predicting}, proving mathematical theorems \citep{gowers2009massively}, or even solving quantum mechanics problems \citep{sorensen2016exploring}. In all of these studies, the actual problem was simplified into an approachable computer game.

Another interesting question in this context is what kind of decision-making process leads to cooperation and which to selfishness? Does this depends on whether the decisions are based on rational calculation or are more intuitive? The decision-making in cooperative games is known to be influenced by time pressure \citep{rand2014social,grujic2018}, yet it is likely to be also influenced by a series of other factors. Since the concept of collaboration is naturally linked to confronting problems in groups, two obvious factors are the \textit{size of the group} confronting the problem, and the \textit{complexity of the problem} itself. The group size as the variable is related to the \textit{peer pressure} exerted by the group on the individual: when in small groups, humans behave differently than when in the crowd \citep{reicher2001psychology}. Actually, the relation between the group size and the benefit to the individual from cooperation has been largely studied \citep{isaac1988group,isaac1994group,barcelo2015g,capraro2015p}. On the other hand, the problem complexity as perceived by an individual can be real or imaginary. Yet regardless of this, it is well-known to objectively influence our decision-making \citep{campbell1988,bystrom1995task}. 

The role of both these variables can be measured via suitable game-theoretic experiment: after decades of successful theoretical developments \citep{helbing2010evolutionary,szolnoki2012wisdom,nosenzo2015cooperation,perc2017statistical} and with the help of modern technology, game theorists can now realize diverse experiments under controlled conditions. They revealed profound aspects of human cooperation in simple models such as Prisoner's dilemma \citep{wong2005,grujic2010social,gracia2012,capraro2014,grujic2014comparative,perc2016phase}, and in  intricate phenomena such as reputation \citep{cuesta2015reputation}, coordination \citep{antonioni2014global}, reward and punishment \citep{rand2009positive}, inter-generational solidarity \citep{hauser2014cooperating}, and for various underlying social networks \citep{gracia2012,rand2014static}.

The question we ask here is: How much error we make when estimating the benefits of collaboration without having the information relevant for that decision available? Does this depend on the size of the social group and the complexity (difficulty) of the problem that our group is trying to solve? Precisely: \textit{given a group of human players solving a problem of specific level of complexity, how much does the empirically found level of collaboration of an average player differs from the level of collaboration that leads to the best attainable score}? This question requires an experiment whose realization is associated with several difficulties. First, how to design a sequence of problems whose levels of complexity are equally perceived by all players? Second, the obtained results should be universal, and not attributable to individual characteristics of any player or the composition of any group of players. Third, we need strong evidence that players do not make their collaborative decisions (exclusively) via rational calculation. Fourth, we need to find the level of collaboration which leads to the best outcome and compare it to the empirically found level of collaboration.

We designed and performed just such an experiment and complement it by the equivalent computer simulation. As we report in what follows, this framework allows to circumvent the above difficulties to a sufficient degree and obtain interpretable findings (we explain later why we believe that players primarily follow their intuition). We found that humans match the optimal strategy only in large groups or when solving hard problems. In contrast, when playing alone, humans massively fail. In agreement with the previous studies, we conclude that in addition to time pressure \citep{rand2012,cone2014,grujic2018,capraro2017does,capraro2018time}, the onset of cooperative behavior is affected by the size of the social group and the complexity of the situation \citep{latane1979many,nowak1990private,gigerenzer1996reasoning,rand2014social}.


\section{The experiment and the simulation}

\paragraph{The experiment with human players.} We recruited 216 participants on a voluntary basis (150 f, 66 m, age $19 - 51$, M = 22.8, SD = 4.53), all students at School of Psychology or School of Engineering of the University of Florence, Italy, who declared themselves native Italian speakers.

Our experiment involves ``solving'' the problems of increasing complexity in groups of increasing size. But how to design a ``problem'' whose level of complexity (difficulty) is objectively seen as equal by all participants \citep{campbell1988,maynard1997}? We first note that we are here not interested in modeling the problem complexity \textit{per se}, but in studying the \textit{effects} that it has on participants (players). With this in mind we represent a ``problem'' as a probability $p$ ($0<p<1$), which models the chance of problem being ``solved''. In other words, we use $p$ as a proxy for problem complexity, so that $p=0$ models a problem impossible to solve (and get reward for it), whereas $p=1$ means problem is automatically solved. For easier orientation we define the level of complexity for a problem as $R=1-p$, so that a problem with $R$ close to 0 is ``easy'', while a problem with $R$ close to 1 is ``hard''. Hence, solving a problem of complexity $R$ means throwing a random number between 0 and 1: if it is smaller (bigger) than $p=1-R$, the problem is solved (not solved). In our experiment players make decisions to collaborate or not when solving the problems at four levels of complexity: $R=0.9, 0.7, 0.5$ and $0.1$ (details follow). Our representation of the problem complexity is crude and unrelated to player's cognitive abilities, but it guarantees that complexity is perceived equally by all players (more details in the Supplement). This is crucial for our experiment, as it establishes the grounds for assuming that players' collaborative decisions are based on a unique perception of the problem.

The experiment was organized in 18 sessions, each involving 12 different participants (players). Each participant was engaged in only one session $(18\times12 = 216)$. One session consisted of two matches played by these 12 players, each match identified by its value of $R$. That is to say that each of the 12 participants only played at two out of the four levels of complexity $R$. Each of the four values of $R$ was used in 9 different matches, played within 9 different sessions, each session attended by 12 different players. By shuffling the participants this way, we can guarantee that the results cannot be attributed to their personal characteristics. In each match the 12 players were divided in groups of equal size $S$. The experiment involved four different schemes of division into groups: $S=1$ (12 groups with one player each), $S=3$ (4 groups with 3 players), $S=6$ (2 groups with 6 players) and $S=12$ (all 12 players in one single group). A match consisted of a sequence of four games played consecutively, one for each division scheme defined by $S$. This implies that each complexity level $R$ used in 9 matches is, in each match, played at all four division schemes.

A game consisted of participants in all groups playing 11 rounds. One round for a player means one decision to cooperate or not with other $S-1$ players in his/her group, facing a problem of complexity $R$. By virtue of each player making a decision in a given round within his/her group, that group is split into cooperators (players that decided to cooperate in that round) and non-cooperators (players that decided otherwise). With a probability $p=1-R$ a player either solves the problem (wins) or fails to solve it (loses). This happens simultaneously in all groups which are part of a division scheme. A round is completed when all players in all groups (cooperators and non-cooperators) had either won or lost. In the next round all players in all groups choose again to cooperate or not, and the process repeats. Once all 11 rounds are completed, a new game starts, in which the 12 players are divided according to the next division scheme $S$, and the scoring is reset to zero. Once all four division schemes are played, the match is completed, and a new match starts at a next complexity level $R$. Once two matches are completed, the session is closed, and a new session starts with 12 new participants invited into the experiment room. Each combination of $R$ and $S$ is played 9 times, each time with different players, making the results not attributable to the compositions of the groups.

Each player has an individual score that is increased in each round via two separate factors: (i) the number $N_c^w$ of cooperators within the group that won in a given round, and (ii) the group bonus $GB$, a separate score related to the entire group, which increases by 1 each time at least one cooperator wins. The scoring per round for a cooperator is:
\begin{equation} 
\label{eq:coopgain}
\mbox{Pay-off}_{\mbox{coop}} = \left(\frac{1 + GB}{S}\right) {N_c^w},
\end{equation}
where $S$ is the number of players in the group (not the number of cooperators). $GB$ increases by 1 each time $N_c^w>0$. Note that the payoff of a cooperator does not depend on whether he/she wins, but just on $N_c^w$ and $GB$.

In contrast to a cooperator, a non-cooperator is allowed to choose his/her own gain $K$ from integers between 1 to 10, with the payoff depending directly on winning. The price for this is decreased winning probability: a non-cooperator wins with probability $p^K$ (decreases exponentially with $K$). The scoring for a non-cooperator who wins is:
\begin{equation} \label{eq:noncoop-win}
\mbox{Pay-off}_{\mbox{non-coop}} (\mbox{win}) = K + \left[\left(\frac{1 + GB}{S}\right) {N_c}^{w}\right], 
\end{equation} 
and is not related to the number of non-cooperators that win. On the other hand, even when loses, a non-cooperator is guaranteed the same payoff as a cooperator:
\begin{equation} \label{eq:noncoop-lose}
\mbox{Pay-off}_{\mbox{non-coop}} (\mbox{lose}) = \left(\frac{1 + GB}{S}\right) {N_c}^{w} = \mbox{Pay-off}_{\mbox{coop}}.
\end{equation}
So non-cooperators earn by exploiting cooperators' efforts, while cooperators have no benefit of non-cooperators' success. $GB$ does not increase when non-cooperators win.

All decisions and scores are stored throughout the experiment. To simplify our analysis, we focus only on the two following variables.
\begin{itemize}
\item \textit{Average probability of cooperation} $C$: the fraction of positive cooperative decisions made by a given player in one game (11 rounds -- 11 decisions),
\item \textit{Agent fitness} $AF$: the final cumulative payoff of a given player after completing one game.
\end{itemize} 
The values of $C$ and $AF$ are measured in all games. For each combination of $R$ and $S$ we thus obtain 9 values of $C$ and $AF$, each for one game played by a different player. We average these 9 values to obtain the universal values of $C$ and $AF$ for the corresponding $R$ and $S$, which are not attributable to characteristics of players/groups.

The entire experiment was based on online communications, since they are easier to record than face-to-face communications and are less sensitive to non-verbal interactions \citep{walther2011theories}, which can distort the results of an experiment (more technical details in the Supplement). Participants also completed a standard socio-demographic and psychological questionnaire. Ethical Commission of the University of Florence was made aware of our experiment, yet under current ethical regulations no explicit approval was required.

\paragraph{The equivalent computer simulation.} Due to its game-theoretic nature, our experiment can be directly translated into a computer simulation, where human players are modeled as evolving agents \citep{guazzini2015}. This was done by letting the (virtual) agents evolve in a competitive environment (similarly to a genetic algorithm). Each agent is defined by a fixed collaboration probability. Unsuccessful agents are replaced by new agents, where collaboration probability is selected at random. This allows us to extract the collaboration probability $C_{\mbox{best}}$ that leads to theoretically best agent fitness $AF_{\mbox{best}}$. We computed $C_{\mbox{best}}$ and $AF_{\mbox{best}}$ by applying our evolutionary algorithm to a large population of evolving agents (other details in the Supplement). This is done separately for each combination of $R$ and $S$, assuming that agents have no knowledge about each others decisions, in accordance with the experimental setting.


\section{Results}

We present our findings by comparing the experimental and simulated values of agent fitness ($AF$ vs. $AF_{\mbox{best}}$) and of average probability of cooperation ($C$ vs. $C_{\mbox{best}}$).
\begin{figure*}[!hbt]
  \centering
  \includegraphics[width=12.1cm]{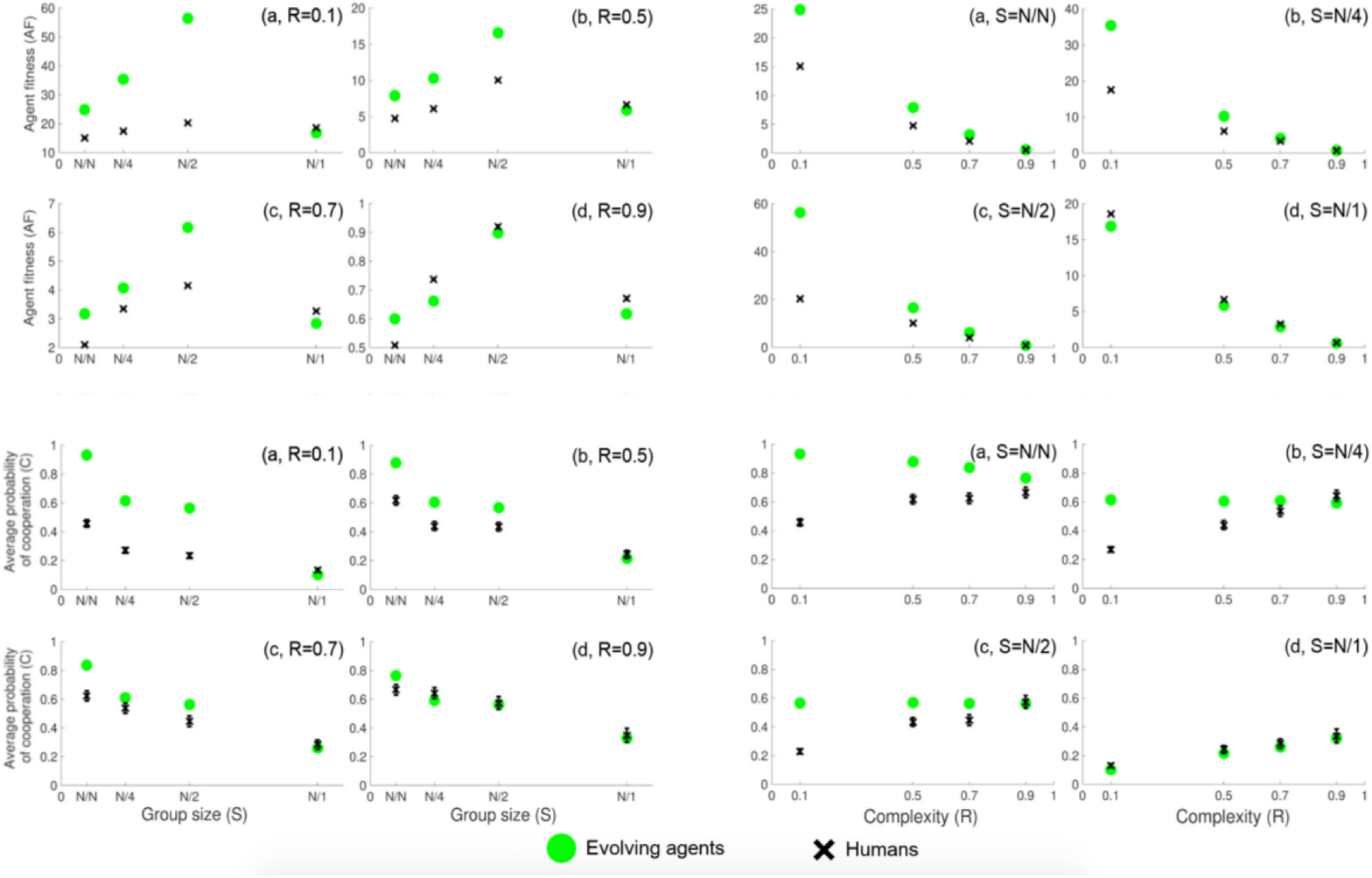}
  \caption{\textbf{Comparison of the experimental and the simulated values of agent fitness and average probability of cooperation.} \textbf{Top two panels}: Comparison of experimental $AF$ (black crosses) and best attainable $AF_{\mbox{best}}$ (green circles). Top left panel: four plots (a)-(d) show the comparison over four values of problem complexity $R$. Within each plot we show the values for four group sizes $S$, where $N/N$ indicates the group composed of a single player, while $N/4$, $N/2$ and $N/1$ respectively indicate the divisions into 4, 2 and 1 group (since experiment and simulation have different total number of players/agents, for easier comparison we use the notation involving number of groups). Top right panel: four plots (a)-(d) show the same comparison, but this time over four group sizes $S$. When dealing with simple problems and when playing in small groups, humans earn much less than they could. This performance improves as problems get harder and/or as humans play in larger groups. Finally, when facing hardest problems or when playing all together in one group, humans earn as much as they could. Humans somewhere appear to earn slightly more than evolving agents: this is an artifact coming from the statistical nature of the simulations (see Supplement). \textbf{Bottom two panels}: Comparison of experimental $C$ (black crosses) and $C_{\mbox{best}}$ leading to best agent fitness (green circles). Bottom left panel: four plots (a)-(d) show the average level of cooperation $C$ for four values of problem complexity $R$. Within each plot we show the values for four group sizes $S$, as in the top panels. Bottom right panel: four plots (a)-(d) show the same comparison, but this time over four group sizes $S$. When in isolation/small groups and/or when confronting simple problems, humans cooperate far less that it would be best for their benefit. As groups become bigger and problems become harder, average human level of collaboration increases. Finally, when all in one group or when facing hardest problems, humans correctly judge the best level of cooperation. All differences between experimental and simulated values in all panels are statistically significant. All values have error bars (very small) that express standard deviations.}
  \label{figure1}
\end{figure*}

\paragraph{Humans systematically earn less than they could and reach their full potential only when confronting the hardest problems or working in the largest group.}  We first examine the differences between $AF$ and $AF_{\mbox{best}}$ in relation to $R$ and $S$. This is also to confirm that our computer simulation did indeed extract the strategy that leads to the best attainable $AF_{\mbox{best}}$. However, $AF$ and $AF_{\mbox{best}}$ cannot be immediately compared, since due to the evolutionary nature of the simulations the two scoring schemes are not normalized in the same way. In order to make an interpretable comparison, we run additional independent simulations as described in the Supplement, and obtain new values of $AF$ and $AF_{\mbox{best}}$, reported in the Fig.\ref{figure1}, top two panels.

Looking at the top left panel in Fig. \ref{figure1} (four plots, a--d), within each plot we show the values for four problem complexities $R$. For simple problems (a) humans earn considerably less than they best can, specially for the group consisting of half of players. Exception is the case of largest group (12 players), where humans actually match the best score. As problems get harder (b and c), the performance of humans is systematically closer to the best performance. The payoffs in largest group are always matching the best payoffs. When facing the hardest problems (d), humans basically earn the maximum possible payoff in groups of any size. In the top right panel in Fig.~\ref{figure1} we show the same values as in the top left panel, but now presenting them via group size $S$. For the smallest group (a) consisting of one player who plays alone, humans largely under-perform, but their performance improves as the problems become harder. As the group size increases (b and c), human performance increasingly matches the best performance -- a trend that is more pronounced for harder problems. When playing in the group comprising all 12 players (d), humans exhibit almost ideal performance regardless of the problems complexity.

As $R$ approaches 1 the values of $AF$ and $AF_{\mbox{best}}$ go to zero, so our result may appear trivial there. But as we show below, the average probability of cooperation $C$ does not go to zero in this limit. We also found that the ratio between $AF_{\mbox{best}}$ and $AF$ is close to 1 in this limit independently of $S$, implying that this is not a mere artifact.

These results indicate a clear correlation between $AF$ and both $R$ and $S$. The performance of humans systematically improves as the problems get harder and as the groups get bigger, eventually reaching the optimum for the hardest problems and largest group. This points to cooperation being more useful in larger groups and for harder problems. In contrast, humans considerably under-perform when solving easy problems and/or when playing in isolation or small groups, which are the situations where cooperation appears not to work to the benefit of players. As we show in what follows, this is related to excessive opportunistic behavior and intuitive  of the benefits of collaboration.

\paragraph{Humans systematically collaborate less than what would lead to the maximal payoff, correctly judging the best collaboration level only when confronting the hardest problems or working in the largest group.} We now examine the differences between $C$ and $C_{\mbox{best}}$ in relation to $R$ and $S$, as reported in the Fig.\ref{figure1}, bottom two panels. No additional normalization is needed here.

In the bottom left panel of Fig.\ref{figure1} (four plots, a--d), within each plot we show the values for four problem complexities $R$. When problems are simple (a), humans largely fail in estimating the optimal collaboration level, and do well only when in the largest group comprising all players. As problems get harder (b and c), humans increasingly better estimate the best collaboration level. In fact, the estimations are systematically better for larger groups. For hardest problems (d), humans correctly guess the best cooperation level regardless of the group size. In the bottom right panel in Fig.~\ref{figure1} we show the same values as in the bottom left panel, but now presenting them via group size $S$. When playing alone (a), humans generally fail to find the best cooperation level, but their estimation improves as the problems become harder. For groups of moderate size (b and c), humans increasingly better guess the optimal cooperation level. In fact, their guesses are improving as the problems get harder, and for hardest problems the estimations are correct. When playing all together in one group (d), humans collaborate at the optimal level for problems of any complexity. Interestingly, humans are never found to collaborate more than the optimal level (which could likewise go against their benefit).

These results are consistent with the findings for agent fitness from the two top panels in Fig.~\ref{figure1}, suggesting a correlation between $C$ and both $R$ and $S$. Humans estimate the optimal collaboration level systematically better as the groups become bigger and the problems become harder. When playing in the largest group or when facing the hardest problems, humans correctly judge the best collaboration level. In contrast, when playing in isolation or when facing easy problems, humans largely fail in their collaborative strategy. These observations explain why the earning of humans are considerably smaller than what they could be, except in cases of largest group and/or hardest problems. This also indicates that collaboration -- whose usefulness we understand via discrepancy between theoretical and real behavior -- is more beneficial in larger groups and for harder problems. These seem to be the conditions when humans attain their full collaborative potential.

We also observe that both humans and evolving agents cooperate more in small groups than in large groups regardless of the problem complexity. Such behavior is known as \textit{social loafing} and is often found in experiments \citep{latane1979many,chidambaram2005out,piezon2008perceptions,abraham2015sociopsychotechnological}. Humans quickly lose interest in solving a problem when inserted in a group of increasing size, but instead count on others to solve it (free riding). On the other hand, both humans and evolving agents cooperate more as the problems get harder. In other words, humans behave opportunistically when facing easy problems, while they resort to collective actions when dealing with complex problems, which holds regardless of the group size. Also interesting is the group involving half of the players (division into 2 equal groups), where humans most drastically under-perform (except in the case of hardest problems). Social loafing here seems to have the strongest negative effect.

\paragraph{Humans systematically generate smaller group bonus than they could, reaching the maximal values only for the hardest problems or when working in the largest group.} One may think that while under-performing for the case of their individual benefit (measured by $AF$), human choose the level of collaboration that actually maximizes the common benefit, measured by group bonus $GB$. This however is not the case: we compare the human $GB$ with the optimal $GB_{\mbox{best}}$ in the Supplement, and find a very similar pattern to what found for $AF$: human level of collaboration generates best $GB_{\mbox{best}}$ only when facing hardest problems or when working in the largest group, and fails under other conditions. Therefore, social collaboration seems to be equally useful to humans for both individual and common benefit.

\paragraph{Psychological variables only weakly correlate with the propensity for collaboration.} We also examined if the psychological variables measured via survey predict the collaborative behavior. We examined standard "Big 5" personality dimensions (Neuroticism, Surgency, Agreebleness, Closeness, and Conscientiousness), Honesty-Humility scale, State-Trait Anxiety Inventory, General Self-Efficacy Scale, and Social Community scale (see Supplement for details). Using standard statistical analysis we show that only a small part of the collaboration variance can be explained via psychological variables. This further confirms that $R$ and $S$ are indeed the only relevant variables affecting the collaborative behavior of players and the findings cannot be attributed to players' personal characteristics.


\section{Discussion}

In sum, we found that in our experiment humans perform best when joined in large groups to solve problems they see as difficult. Under these conditions humans most accurately estimate the level of collaboration that will maximize their benefit. In contrast, the same approach fails when humans act in small groups or face challenges they perceive as easy. These findings are robust to players' and groups' characteristics including personal traits, and appear to depend exclusively on the two game-theoretic variables in our setup: problem complexity and group size. 

We interpret these findings as a result of players' intuitive (heuristic) decision-making. In fact, recent and not-so-recent results in cognitive psychology consistently show that humans are not always rational \citep{simon1972theories,kahneman2002representativeness}. This was found in a series of experiments, indicating that our processing of information can follow either a fast and economic \textit{heuristic} path or a slower and more resource-demanding \textit{rational} path \citep{tversky1974judgment,gigerenzer1996reasoning,shah2008heuristics}. The choice between these two paths depends on many factors, such as emotional involvement with the situation \citep{slovic2007affect,jaspersen2017influence} or time pressure to react in a situation \citep{rand2016cooperation,bobadilla2018fast,grujic2018}. Specifically, our (often) irrational drive for cooperation was recently articulated as \textit{Social Heuristics Hypothesis} \citep{rand2014social}: It posits that through life experience we devise simple rules-of-thumb that enable us to instinctively adopt ``socially best'' choices, as opposed to a more rational, slow and often selfish consideration. And indeed, when the decisions are taken under time pressure, intuitive reasoning leads to more cooperation in social dilemma games \citep{rand2014reflection,capraro2017deliberation} and in dictator games \citep{rand2016social}. In contrast, with no rush to decide, humans perform a more accurate evaluation of their individual interests and cooperate less \citep{belloc2018social}. We appear to internalize and learn \citep{capraro2015social} simple intuitive strategies that are advantageous in our routine lives and ``mistakenly'' transplant them into less typical situations, where we would be (individually) better off with a more rational approach. Guiding ourselves on heuristics generally works to our benefit, but under specific conditions can also work against our benefit.

Coming back to our experiment, we cannot conclusively prove the absence of deliberative rational thinking in players' decision-making. However, below we discuss four independent lines of evidence pointing to this. 

First, humans are known to generally apply intuition when limited information suggests that it could be the best strategy. Because time pressure regulates the speed by which subjects have to process the available information, so it can be seen as a proxy for the availability of information. Along the same lines, time pressure can also be seen as a way to manipulate the perceived complexity of the problem: solving a fixed problem under varying time constraints generates varying perception of the difficulty of that problem. In our experiment we simply use a different way to regulate the perceived complexity of the problem -- we manipulate the probability for a player to win in a given round. This is a good proxy for a regulable problem complexity, much like the time pressure. So from this point of view we believe that our experimental setup is not that different from the setups in earlier studies on SHH. 

Second, in our scoring scheme cooperators contribute not only to the common payoff, but also to the payoff of non-cooperators, who can further increase their individual payoffs by striving for a higher reward at the price of solving the harder task alone. In these conditions we expect that the cooperative decisions are taken via same heuristics as in the real life: by sharing a piece of information a player can increase the benefit for those who exploit this wisdom, whereas by choosing to work alone, a player can profit of the collective knowledge (articulated as group bonus) without contributing to it. As a side effect, this complicates the scoring scheme and makes it impossible for players to calculate any payoffs beforehand. In addition, players cannot infer the strategies of their peers since they have no access to their history of cooperative decisions. So, not having much to go on, we expect that players mainly followed their instincts from round to round. In other words, limited information about other players should preclude the application of any deliberative reasoning.

Third, when the community is divided into groups (that could also be competing among them), rational reasoning would make the collaboration level depend primarily on the group one belongs to and on the total number of groups, rather than on the problem complexity. In fact, we such behavior is found for evolving agents, whose cooperation deceases with increase of $R$, except in the case of the largest group ($S=N/1$, cf. Fig.\ref{figure1}), where the cooperation actually increases with increase of $R$. Humans instead display a different pattern: they increase their cooperation with increase of $R$ for all group sizes, including the largest group, which is hard to ascribe to any systematic rational reasoning. 

Fourth, the typical reaction time of players did not change significantly over different rounds and games. In fact, when players reason rationally they are known to be slower to make decisions \citep{rand2014social,grujic2018}. Therefore, rational thinking would in our case make the reaction time vary or even increase as the game unfolds, since players ``learn'' how to play in their best interest. However, as we show in the Supplement, players' reaction times are not correlated with their propensity to collaborate, meaning that no learning effects are present. In other words, players in all rounds seem to make their decisions on the same ground as in the first round. But if players are not learning as the game unfolds, it is reasonable to hypothesize that they rely on a previously learned strategy. And all heuristics, including the social heuristics, are well-known to be socially learned via life experience \citep{gigerenzer1999fast,chaiken1989heuristic}. So, while we cannot determine conclusively that players relied \textit{exclusively} on heuristics and not on rational thinking, above body of evidence strongly points to this interpretation.

Game-theoretic nature of our experiment allowed us to use computer simulation to exactly reproduce the experimental conditions. But how realistic is to approximate a human player as an evolving agent defined by the constant cooperation probability? Human players do not make their decisions ``randomly'', but based on some strategy or feeling, which makes this approximation crude. However, as we explained above, cooperative decisions taken in separate rounds cannot be statistically distinguished across players nor across rounds. Hence, as far as the external observer (experimenter) is concerned, a player can be well characterized by a single number, which expresses how likely is that he/she will cooperate in any given round. This also indicates that even more a sophisticated simulation of our experiment is not likely to reproduce more accurate estimation of the best strategy.

To what degree are these findings dependent on the choice of the game, especially in relation to the opposition between individual and common benefit? We designed our game to best reproduce a cognitive scenario that would force the players to act as much as possible on intuition. A game with a strict opposition between individual and common benefit would place a ``simpler'' dilemma in front of the players, making it harder for us to exclude the rational thinking. In fact, replication of our experiment using other games, possibly with varying balance of individual and common benefit, is the prime direction of future work. One could also examine the influence of a combination of several factors on the usefulness of collaboration heuristics, for example, a combination of time pressure, group size and problem complexity. Other directions include replication of the experiment with larger groups and more realistic (ecological) problems. While the former is generally doable, the latter is associated with already discussed difficulty (operationalization of problem complexity). Actually, we admit that the ultimate interpretation and the value of our results might be depended on our definition of problem complexity. We warmly welcome any further work where players would be solving actual problems of gradually increasing complexity.  
In conclusion, we argue that humans have evolved to solve demanding challenges best when joining forces in large groups, which is in agreement with the existing studies \cite{prelec2017solution,tausczik2017size,huang2018does}. This explains why errors in our collaboration judgment are made more often and more drastically in smaller groups and when dealing with easier problems. At a certain point in our evolutionary path \cite{bowles2006group} communities whose members have been able to cooperate in the appropriate measure have thrived, as opposed to those communities whose members collaborated too much or not enough. Actually, there are numerous hypotheses on the evolutionary mechanisms perpetuating the human cooperative behavior: group/kin selection, direct/indirect reciprocity, etc. \citep{bowles2004evolution,nowak2005evolution,nowak2006,rand2009direct}. We here argue that an additional factor has been the need to find the level of collaboration that would best benefit both the community and the individual. \\[0.3cm]


\end{document}